\documentclass[traditabstract]{aa}

\usepackage{natbib}
\usepackage{lscape}
\usepackage[section]{placeins}
\usepackage{wasysym}
\usepackage{graphics}
\usepackage{graphicx}
\usepackage{txfonts}
\usepackage{color}
\usepackage{times}

\usepackage{hyperref}
\hypersetup{
    bookmarks=true,         
    unicode=false,          
    pdftoolbar=true,        
    pdfmenubar=true,        
    pdffitwindow=true,      
    pdftitle={},    
    pdfauthor={Gomes \& Papaderos},     
    pdfsubject={RemoveYoung: A tool for the removal of the young stellar component in galaxies within an adjustable age cutoff},      
    pdfnewwindow=true,      
    pdfkeywords={keywords}, 
    colorlinks=true,        
    linkcolor=blue,         
    citecolor=blue,         
    filecolor=blue,         
    urlcolor=blue           
}

\usepackage{mathtools}
\usepackage[T1]{fontenc}

\newfont{\xnlx}{cmssdc10 scaled 800}
\newfont{\nlx}{cmssdc10 scaled 900}
\newfont{\vcap}{cmssdc10 scaled 1000}
\newfont{\lcap}{cmssdc10 scaled 1100}
\newfont{\icap}{cmssdc10 scaled 1600}
\newfont{\mcap}{cmssdc10 scaled 2000}
\newcommand{\rem}[1]{\nlx {#1}\normalfont}

\newcommand{\sbb}{mag/$\sq\arcsec$}

\def\rr{{\sl R}$^{\star}$}
\def\ha{H$\alpha$}

\def\sstar{$\Sigma_{\star}$}
\def\lo3hb{$\log$([O\,{\sc iii}]/H$\beta$)}
\def\ln2ha{$\log$([N\,{\sc ii}]/H$\alpha$)}
\def\tauha{$\tau$}
\def\tauha_ext{$\tau$}

\def\rr{{\sl R}$^{\star}$}

\def\eqan{\begin{equation}}
\def\eqen{\end{equation}}

\def\h1{\ion{H}{i}}
\def\h2{\ion{H}{ii}}

\def\rr{{\sl R}$^{\star}$}

\def\ha{H$\alpha$}

\def\sstar{$\Sigma_{\star}$}

\def\o5007{[O {\sc iii}] $\lambda$5007}

\def\n2ha{[N\,{\sc ii}]/H$\alpha$}
\def\ln2ha{$\log$([N\,{\sc ii}]/H$\alpha$)}
\def\tn2ha{[N\,{\sc ii}]${\scriptstyle 6584}$/H$\alpha$}
\def\tln2ha{$\log$([N\,{\sc ii}]${\scriptstyle 6584}$/H$\alpha$)}
\def\o3hb{[O\,{\sc iii}\,]/H$\beta$}
\def\lo3hb{$\log$([O\,{\sc iii}]/H$\beta$)}
\def\to3hb{[O\,{\sc iii}]${\scriptstyle 5007}$/H$\beta$}
\def\tlo3hb{$\log$([O\,{\sc iii}]${\scriptstyle 5007}$/H$\beta$)}
\def\e16{$10^{-16}~{\rm erg\,s^{-1}\,cm^{-2}}$}
\def\e17{$10^{-17}~{\rm erg\,s^{-1}\,cm^{-2}}$}
\def\tauha{$\tau$}
\def\tauha_ext{$\tau$}
\def\ha{H$\alpha$}

\def\sstar{$\Sigma_{\star}$}
\def\tauha{$\tau$}
\def\tauha_ext{$\tau$}

\def\mstar{${\cal M}_{\star}$}
\def\ml{${\cal M/L}$}
\def\sstar{$\Sigma_{\star}$}

\def\porto3d{\sc Porto3D\rm}
\def\p3d{\sc Porto3D\rm}
\def\P3D{\sc Porto3D\rm}
\def\SL{{\sc Starlight}\rm}
\newcommand\btab[5]{\begin{table}[#1]\label{#3}{\parbox{#4}{\caption{#2}}\rule[-0.5ex]{0cm}{0.5ex} }
\begin{tabular*}{#4}{#5} \label{#3} }

\newcommand{\etab}[3]{
\end{tabular*}

\vspace*{#1}
\begin{flushleft}
\parbox{#2}{\footnotesize #3}
\end{flushleft}
\end{table} }
\newcommand{\setab}[3]{
\end{tabular*}
\end{footnotesize}

\vspace*{#1}
\begin{flushleft}
\parbox{#2}{\footnotesize #3}
\end{flushleft}
\end{table} }

\def\eqan{\begin{equation}}
\def\eqen{\end{equation}}

\def\h1{\ion{H}{i}}
\def\h2{\ion{H}{ii}}

\newcommand{\zsun}{$Z_\odot$}
\def\rr{{\sl R}$^{\star}$}
\def\P25{{\sl R}$_{\rm SF}$}
\def\E25{{\sl R}$_{\rm host}$}

\def\m5{${\cal M}_{\star,{\rm 5\,Gyr}}$}

\def\D4000{$D_{4000}$}

\def\uflux{erg\,s$^{-1}$\,cm$^{-2}$}

\def\ha{H$\alpha$}

\def\e16{$10^{-16}~{\rm erg\,s^{-1}\,cm^{-2}}$}
\def\e17{$10^{-17}~{\rm erg\,s^{-1}\,cm^{-2}}$}
\def\tcut{{\sl t}$_{\rm cut}$}
\newcommand{\PutLabel}[3]{\put(#1,#2){#3}}
\newcommand {\aga} {\ {\raise-.5ex\hbox{$\buildrel>\over\sim$}}\ }
\newcommand {\ala} {\ {\raise-.5ex\hbox{$\buildrel<\over\sim$}}\ } 
\def\ry{${\cal RY}$}
\def\sps{\nlx sps\rm}

\begin{document}

%
\titlerunning{RemoveYoung: A tool for the removal of the young stellar component in galaxies}
\authorrunning{J.M. Gomes \& P. Papaderos}
\title{RemoveYoung: A tool for the removal of the young stellar component in galaxies within an adjustable age cutoff
}

\author{
J.M. Gomes\inst{\ref{inst1}}
\and P. Papaderos\inst{\ref{inst1}}
}

\institute{
Instituto de Astrofísica e Ci\^encias do Espa\c{c}o, Universidade do Porto,
Centro de Astrof{\'\i}sica da Universidade do Porto, Rua das Estrelas, 4150-762 Porto, 
Portugal\label{inst1}
}
\date{Received ?? / Accepted ??}
\abstract{The optical morphology of galaxies holds the cumulative record of their assembly history, 
and techniques for its quantitative characterization offer a promising avenue toward understanding 
galaxy formation and evolution. However, the morphology of star-forming galaxies is generally dictated 
by the youngest stellar component, which can readily overshine faint structural/morphological 
features in the older underlying stellar background (e.g., relics from recent minor mergers) that could 
hold important insights into the galaxy build-up process.
Stripping off galaxy images from the emission from stellar populations younger than an adjustable age 
cutoff \tcut\ can therefore provide a valuable tool in extragalactic research.
RemoveYoung (\ry), a publicly available tool that is presented here, exploits the combined power of 
integral field spectroscopy (IFS) and spectral population synthesis (\sps) toward this goal.
Two-dimensional (2D) post-processing of \sps\ models to IFS data cubes with \ry\ permits computation of the spectral energy, 
surface brightness, and stellar surface density distribution of stellar populations older than a user-defined \tcut. This
suggests a variety of applications of star-forming galaxies, such as
interacting or merging galaxy pairs and lower mass 
starburst galaxies near and far; these include blue compact and tidal dwarf
galaxies.
} 
\maketitle

\section{Introduction \label{intro}}
How the assembly history of galaxies is imprinted on their present-day
optical morphology is one of the most tantalizing enigmas in
extragalactic research.  Morphology holds the cumulative record of
complex and highly interlinked processes operating on different
temporal and spatial scales across cosmic time, such as quasi-monolithic gas collapse into classical bulges and galaxy
spheroids, gentle gas dissipation into galactic disks, hierarchical
growth via minor and major mergers, and environmentally modulated star
formation (SF) in galaxy pairs/groups and clusters \citep[see][for a
  review]{KormendyKennicutt04}.  Various quantitative morphology
indicators \citep[\rem{QMIs}; see][for a review]{Conselice14} have
been proposed in recent decades and extensively employed for the
characterization of large extragalactic probes in the quest of
elucidating the link between morphology, structure, intrinsic physical
properties (e.g., stellar mass \mstar\ and surface mass density
\sstar, metallicity) and the evolutionary and dynamical status of galaxies.

A first-order approximation in these studies is that the optical
surface brightness $\mu$ traces the \sstar, which essentially presumes
that the stellar mass-to-light ratio (\ml) spans a rather narrow range
of values across the galaxy extent.  Whereas this assumption is
certainly justified for quiescent galaxies or systems with a smooth
star formation history (SFH), it cannot be maintained for systems
exhibiting a high specific star formation rate (sSFR), such as
isolated and interacting starburst galaxies.  The optical appearance
of such galaxies primarily reflects the 2D distribution of the young
($\la$100 Myr) stellar component, which owing to its very low
\ml\ throughout overshines the older underlying
stellar background that is dynamically dominant.

As an example, in a typical blue compact dwarf (BCD) galaxy the
centrally confined starburst component dictates the observed
line-of-sight intensity and contributes 50\% to 80\% of the total
optical emission \citep[][]{P96a,C01,GdPM2005}.  In BCDs and their
higher-$z$ analogs \citep[e.g., {\sl green
    peas};][]{Cardamone2009,Izotov2011,Amorin2012,Amorin2015} SF
typically dominates down to $\mu \simeq$ 24.5-25.0 $B$ \sbb,
i.e., almost out to the Holmberg radius.

This strong disparity between $\mu$ and \sstar\ in star-forming
galaxies has a twofold effect: First, starburst emission can strongly
impact light concentration indices
\citep[e.g.,][]{Morgan1958,Abraham1996} commonly used in \rem{QMI}
studies.  In a typical BCD, for instance, the ignition of a central
starburst shrinks the optical effective radius by $\sim$70\%
\citep{P06} and can mimick a S\'ersic profile with a high ($\eta$=2-4)
exponent \citep[][see also Bergvall \& \"Ostlin 2002]{P96a},
eventually leading to its erroneous classification as an early-type
galaxy.  Secondly, a tiny substrate of luminous young stars can
readily masquerade fainter morphological features that potentially
hold key insights into the recent assembly history and structural
properties of galaxies (e.g., shells and ripples as relics from minor
mergers; cf.  Schweizer \& Seitzer (1988) or intrinsically faint
bars).
%
\begin{figure*}
\begin{picture}(18.4,6.7)
\put(2.0,0.0){\includegraphics[width=12cm, angle=0, viewport=80 470 460 800]{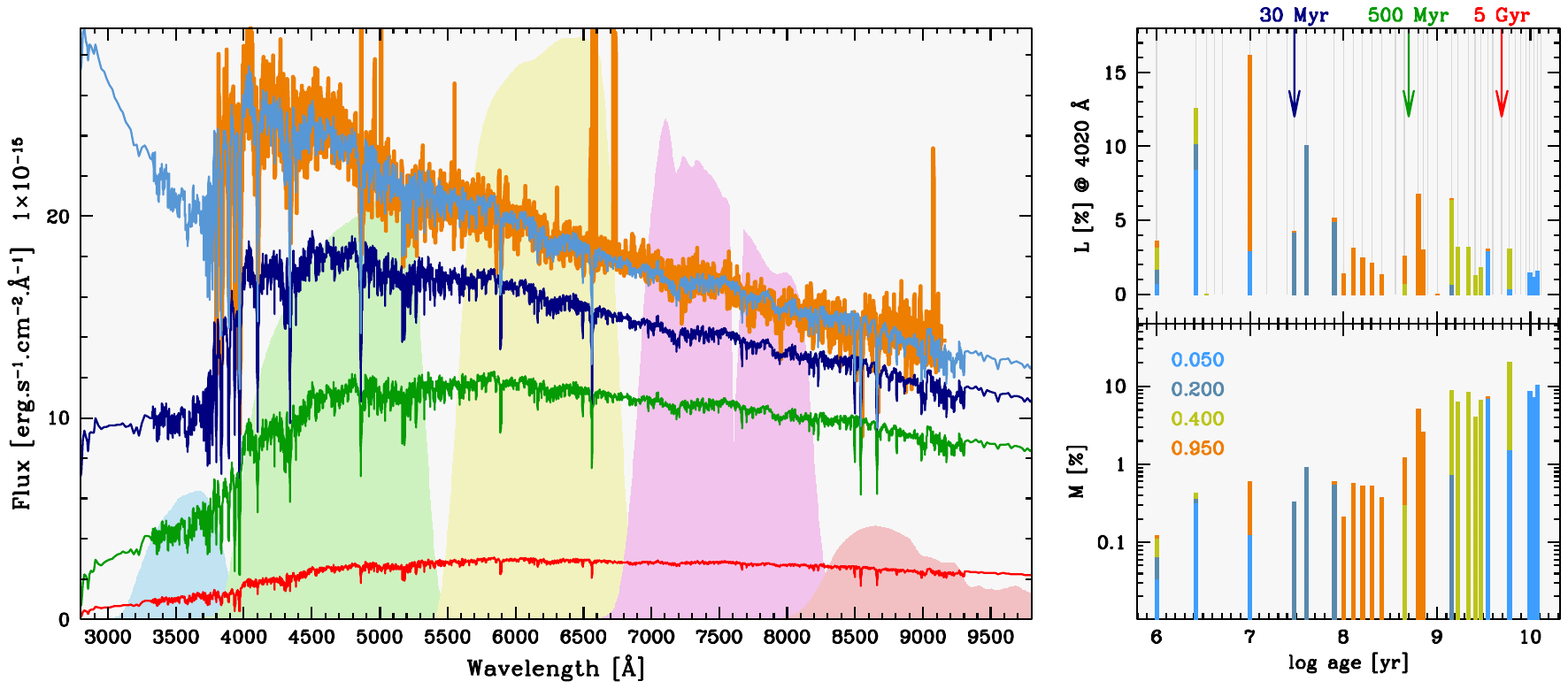}}
\PutLabel{10.6}{6.2}{\mcap a}
\PutLabel{13.0}{6.2}{\mcap b}
\PutLabel{13.80}{3.0}{\mcap c}
\end{picture}
\caption[]{ \rem{a)} Example of the application of \ry\ to the
  best-fitting synthetic SED (light blue) obtained with \SL\ for a
  SDSS-DR7 \citep{yor00} spectrum (orange) for three \tcut\ values: 30
  Myr (blue), 0.5 Gyr (green), and 5 Gyr (red).  The transmission
  curves of the five SDSS filters $u$, $g$, $r$, $i,$ and $z$ are
  depicted with shaded areas.  Panels \rem{b} and \rem{c} show, respectively, the luminosity ($x_j$) and mass ($\mu_j$) contribution
  (\%) of individual SSPs (1 \dots $j$) to the best-fitting population
  vector.  The light-gray vertical lines in the upper-right panel show
  the ages available in the SSP library for four metallicities
  (between 0.05 and 0.95 \zsun) and vertical arrows depict the applied
  \tcut\ values. From panels \rem{b} and \rem{c} it is apparent that,
  according to the best-fitting population vector, SSPs younger than
  30 Myr contribute less than 2\% of \mstar\ but nearly 35\% of the
  observed intensity at the normalization wavelength of 4020 \AA. As a
  result, suppression of these young SSPs results in a significant
  dimming by 0.56, 0.27, 0.22, and 0.19 mag in the SDSS $u$, $g$, $r,$
  and $i$ bands, respectively.
}
\label{fig:RY-spectrum}
\end{figure*}

In light of such considerations, a technique permitting
suppression from galaxy images of the luminosity contribution from
stars younger than an adjustable age cutoff \tcut\ appears especially
useful. Such a tool would not only be valuable to studies of
\rem{QMI}s, but also to those exploring SF patterns back to a
well-defined time interval (e.g., since the infall of a galaxy onto a
cluster). This task is obviously out of reach with standard
techniques, such as near-infrared (NIR) imaging
\citep[e.g.,][]{Noeske03} or state-of-the-art modeling of the observed
spectral energy distribution \citep[SED; e.g.,][]{Wuyts2012}.

In this article, we present a publicly available\footnote{A thoroughly
  documented version of the code is available at {\tt
    http://www.spectralsynthesis.org}.} tool, RemoveYoung (\ry), which
exploits the combined power of integral field spectroscopy (IFS) and spectral
population synthesis (\sps) with the goal of stripping off galaxy images from
the luminosity contribution from stellar populations younger than a
user-defined age cutoff.  The concept is outlined in Sect.~\ref{RY-meth} and
illustrated through its application to IFS data in Sect.~\ref{RY-IFS}, with a
discussion of its potential merits to various subjects of extragalactic
research following in Sect.~\ref{disc}.

\section{Concept and realization of \ry\ \label{RY-meth}}
The concept of \ry\ essentially consists in the elimination from a
synthetic SED of the contribution from stellar populations younger
than an adjustable age cutoff \tcut\ and reconstruction of the
residual SED of the older stellar component.  The input SED is
generally obtained through \sps\ modeling of an observed spectrum as a
linear combination of simple stellar population (SSP) spectra,
each fully characterizing the SED of an instantaneously formed stellar
population of a given metallicity and age.
The best-fitting population vector (PV) from a \sps\ model $M_\lambda$
encapsulates the contribution of individual SSPs along with the
derived intrinsic extinction and stellar velocity dispersion.
Mathematically, \ry\ acts essentially as a `signal post-processing'
tool for the recovered PV.  The best-fitting model $M_\lambda$ is
multiplied by the Heaviside unit step
function\footnote{$H(t-\textrm{\tcut})=\int_{-\infty}^{t-\textrm{\tcut}}
  \delta(s)\ ds = 0$ for $t<$~\tcut\ and $1$ for $t\ge$~\tcut.}
$H(t-\textrm{\tcut}),$ permitting suppression from the SED of the
contribution from stellar populations that are younger than an age cutoff
\tcut\ to the SED

\noindent \ry$_\lambda$\ $ = M_\lambda\ H(t-\textrm{\tcut}) =
M_{\lambda_0} \left[ \sum\limits_{j=1}^{N_\star}
  x_j~b_{j,\lambda}~r_\lambda \right] \otimes
G(v_\star,\sigma_\star)\ H(t-\textrm{\tcut}),$

\noindent where $b_{j,\lambda} \equiv L_\lambda^{SSP}(t_j,Z_j)
/L_{\lambda_0}^{SSP}(t_j,Z_j)$ is the spectrum of the $j^{\rm th}$ SSP
normalized at $\lambda_0$, $r_\lambda \equiv 10^{-0.4(A_\lambda -
  A_{\lambda_0})}$ is the reddening term parametrized by the stellar
extinction $A_\lambda$ at $\lambda$, {\boldmath
  $x$}$=$$(x_1,...,x_{N_\star})$ is the population vector,
$M_{\lambda_0}$ is the synthetic flux at the normalization wavelength,
$N_\star$ is the total number of SSPs, and $G(v_\star,\sigma_\star)$
is the line-of-sight stellar velocity distribution modeled as a
Gaussian centered at velocity $v_\star$ and broadened by
$\sigma_\star$.

In the following, we use for the sake of demonstration
best-fitting PVs from the \sps\ code \SL\ \citep{cid05}. We note,
however, that \ry\ can be applied to the output from any other
\sps\ code.  Figure~\ref{fig:RY-spectrum}a illustrates the
post-processing with \ry\ of a spectral model that is based on a
library of 152 SSPs from \citet{bru03} spanning an age between 1 Myr
and 13 Gyr for four metallicities (0.05, 0.2, 0.4, and 0.95 \zsun).
Panels \rem{b}\&\rem{c} show, respectively, the light and mass
contribution (\%) of the library SSPs composing the PV.  The structure
of the latter, as a linear superposition of SSPs, facilitates
straightforward reconstruction of the synthetic SED of the stellar
component older than any age cutoff, as illustrated in panel \rem{a}
for three \tcut\ values (30 Myr, 0.5 Gyr, and 5 Gyr).

Also, \ry\ computes and exports several other quantities, such as the apparent
magnitude of the stellar component with age $t$ $\lessgtr$ \tcut\ in several
standard broadband filters (Johnson-Cousins, Bessel, SDSS, 2MASS and others),
with the provision of correction for intrinsic extinction.
Further important features of \ry\ are computation of UV-through-NIR broadband
magnitudes from the best-fitting SED and hybrid observed+synthetic
magnitudes through substitution of spectral intervals of a filter transmission
curve eventually not covered by observations by the modeled spectrum; for
example, estimation of the SDSS $i$ band magnitude from a low-resolution
spectrum from the CALIFA IFS survey spanning a spectral range between 3745
\AA\ and truncated at 7300 \AA.  Besides the magnitudes and SEDs of stellar
populations with $t\lessgtr$\tcut, \ry\ also exports \mstar, both prior to and
after correction for the mass fraction returned into the ISM.
%
\subsection{Two-dimensional application of \ry\ to IFS data cubes\label{RY-IFS}}
\begin{figure*}
\begin{picture}(18.4,2.6)
\put(0,0.1){\includegraphics[height=6.0cm, angle=0, scale=0.46]{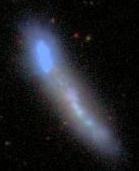}}
\put(2.4,0){\includegraphics[height=6.0cm, angle=0, scale=0.48]{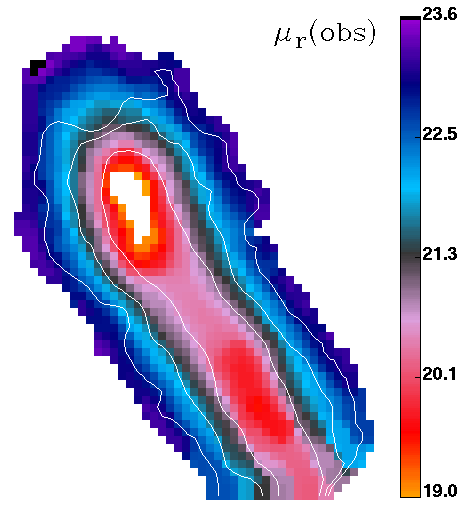}}
\put(5.1,0){\includegraphics[height=6.0cm, angle=0, scale=0.48]{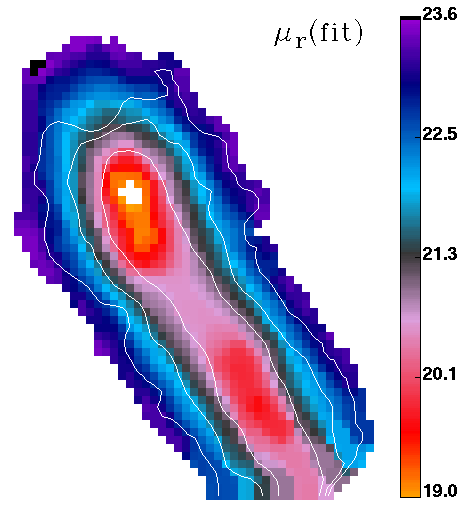}}
\put(7.8,0){\includegraphics[height=6.0cm, angle=0, scale=0.48]{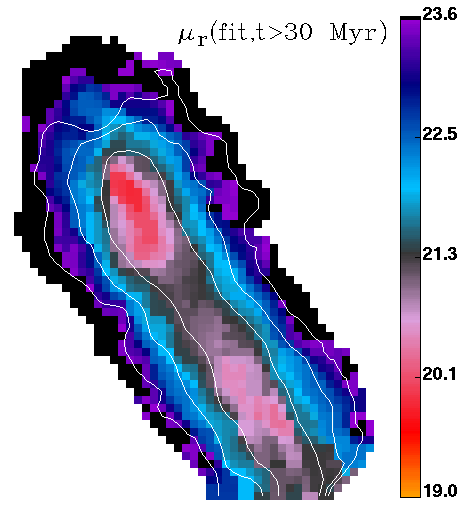}}
\put(10.5,0){\includegraphics[height=6.0cm, angle=0, scale=0.48]{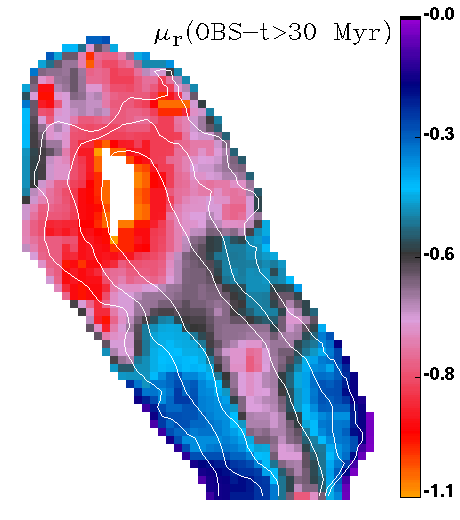}}
\put(13.2,0){\includegraphics[height=6.0cm, angle=0, scale=0.48]{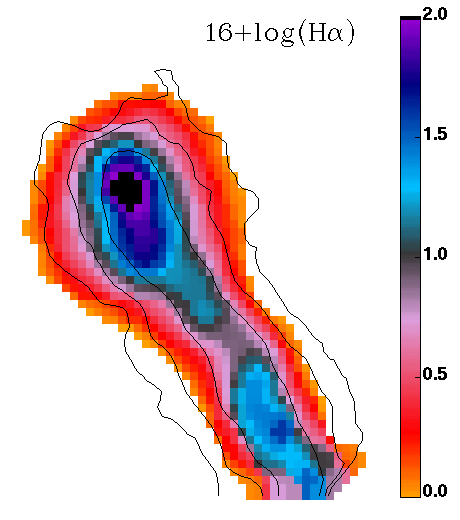}}
\put(15.9,0){\includegraphics[height=6.0cm, angle=0, scale=0.48]{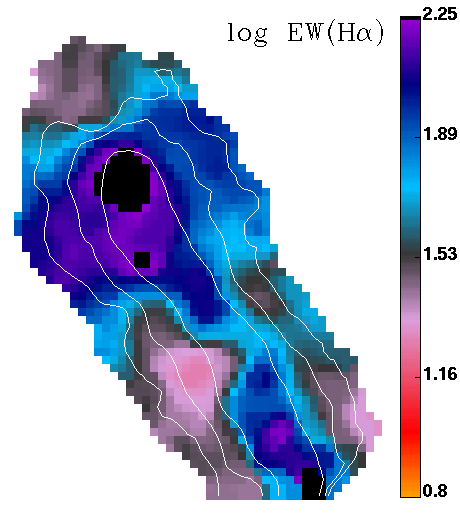}}
\PutLabel{1.4}{2.5}{\color{white}\nlx SDSS}
\PutLabel{0.2}{0.2}{\color{white}\icap a}
\PutLabel{2.6}{0.15}{\color{black}\icap b}
\PutLabel{5.3}{0.15}{\color{black}\icap c}
\PutLabel{8.0}{0.15}{\color{black}\icap d}
\PutLabel{10.7}{0.15}{\color{black}\icap e}
\PutLabel{13.4}{0.15}{\color{black}\icap f}
\PutLabel{16.1}{0.15}{\color{black}\icap g}
\end{picture}
\caption[]{Two-dimensional application of \ry\ on CALIFA IFS data for the
  star-forming galaxy \object{NGC 3991}. \rem{a:} true-color SDSS
  image composite; \rem{b-d:} surface brightness maps (\sbb) computed
  by spaxel-by-spaxel convolution of the $r$-band filter transmission
  curve with the observed spectrum ($\mu_{\rm obs}$; panel \rem{b}),
  the best-fitting stellar SED to the observed spectrum
($\mu_{\rm fit}$; panel \rem{c}),
  and after removal with \ry\ of stellar and nebular emission
  associated with ongoing or recent ($\leq$30 Myr) star formation
  ($\mu$(OBS-30 Myr); panel \rem{d}). Subtraction of the latter from
  $\mu_{\rm obs}$ yields the $\mu$ enhancement (in mag) owing to the
  recent SF (panel \rem{e}), which shows a close spatial correlation
  with the \ha\ flux (in log\ $10^{-16}$ \uflux; panel \rem{f}) and
  \ha\ equivalent width (in \AA; panel \rem{g}).}
\label{RY-n3991}
\end{figure*}

Two-dimensional applications of \ry\ for stripping off the young stellar component
from IFS data cubes are greatly aided by the structured format in which the computed quantities are stored.
To demonstrate the 2D functionality of the code, we apply it next on two nearby SF galaxies
from the second public data release \citep{GarciaBenito15} of the
CALIFA galaxy survey \citep{Sanchez2012}.  Both systems were initially
processed with our automated IFS pipeline \p3d\ \citep{P13,G16a}, which
allows for spectral modeling with \SL\ and derivation of emission-line
fluxes, equivalent widths (EWs), and kinematics.  Fits were computed
spaxel-by-spaxel between 4000 \AA\ and 6800 \AA\ using the same SSP
library as in Sect. \ref{RY-meth}.
\begin{figure*}
\begin{picture}(18.4,2.4)
\put(-0.1,0){\includegraphics[height=2.4cm, angle=0]{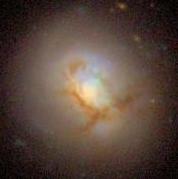}}\\
\put(2.4,0){\includegraphics[height=2.4cm, angle=0]{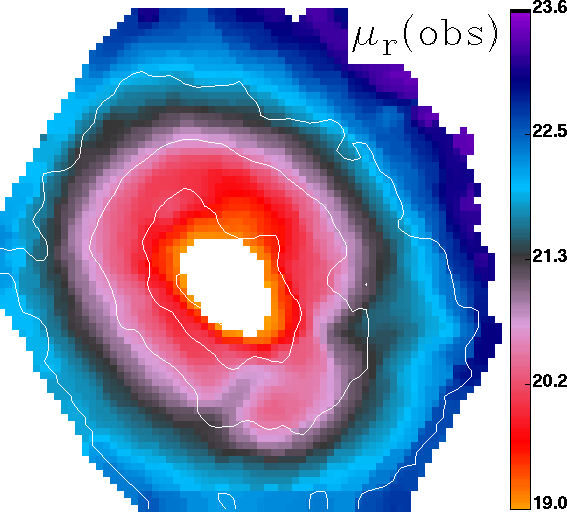}}
\put(5.1,0){\includegraphics[height=2.4cm, angle=0]{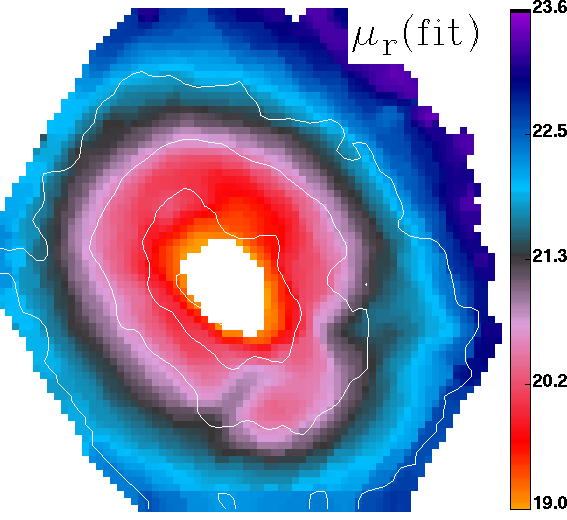}}
\put(7.8,0){\includegraphics[height=2.4cm, angle=0]{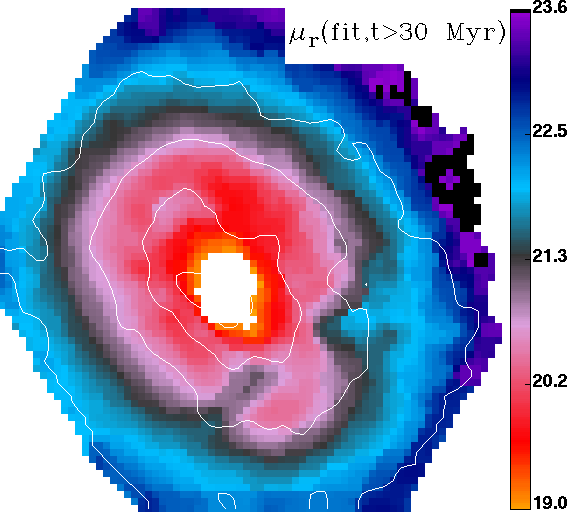}}
\put(10.5,0){\includegraphics[height=2.4cm, angle=0]{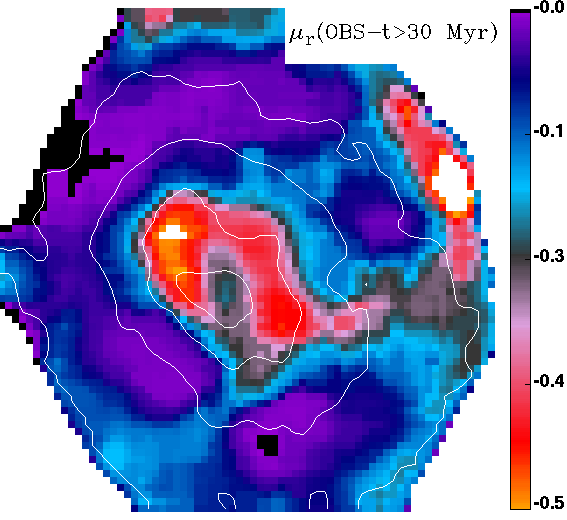}}
\put(13.2,0){\includegraphics[height=2.4cm, angle=0]{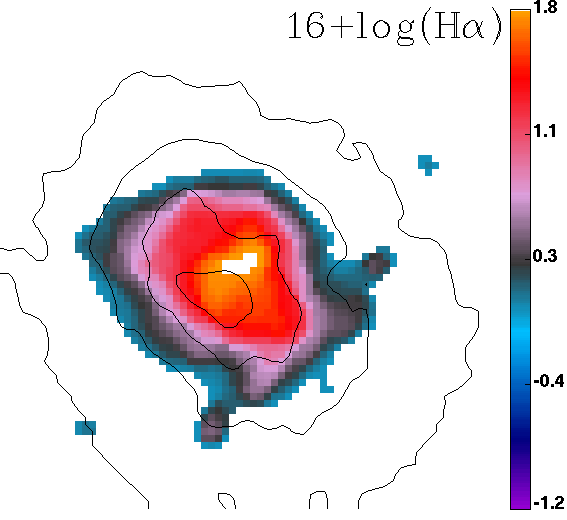}}
\put(15.9,0){\includegraphics[height=2.4cm, angle=0]{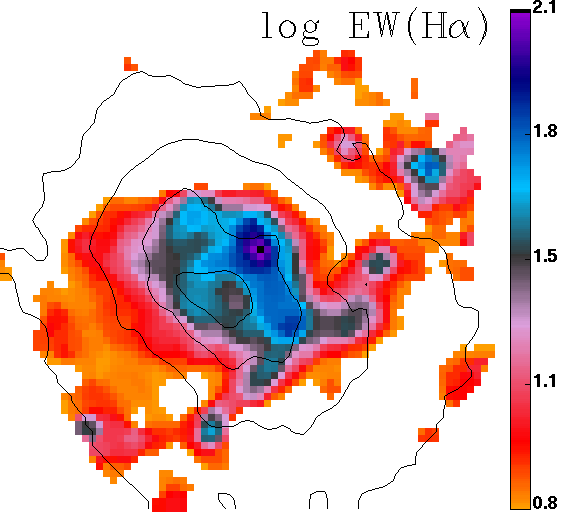}}
\PutLabel{1.4}{2.0}{\color{white}\nlx SDSS}
\PutLabel{0.1}{0.2}{\color{white}\icap a}
\PutLabel{2.4}{0.12}{\color{black}\icap b}
\PutLabel{5.1}{0.12}{\color{black}\icap c}
\PutLabel{7.8}{0.12}{\color{black}\icap d}
\PutLabel{10.5}{0.12}{\color{black}\icap e}
\PutLabel{13.2}{0.12}{\color{black}\icap f}
\PutLabel{15.9}{0.12}{\color{black}\icap g}
\end{picture}
\caption[]{Application of \ry\ on CALIFA IFS data for the luminous BCD galaxy \object{NGC 7625}. The panels have the same meaning 
as in Fig.~\ref{RY-n3991}.
}
\label{RY-n7625}
\end{figure*}

\begin{figure}
\begin{picture}(8.4,4.6)
\put(-1,-0.7){\includegraphics[height=6.3cm, angle=0]{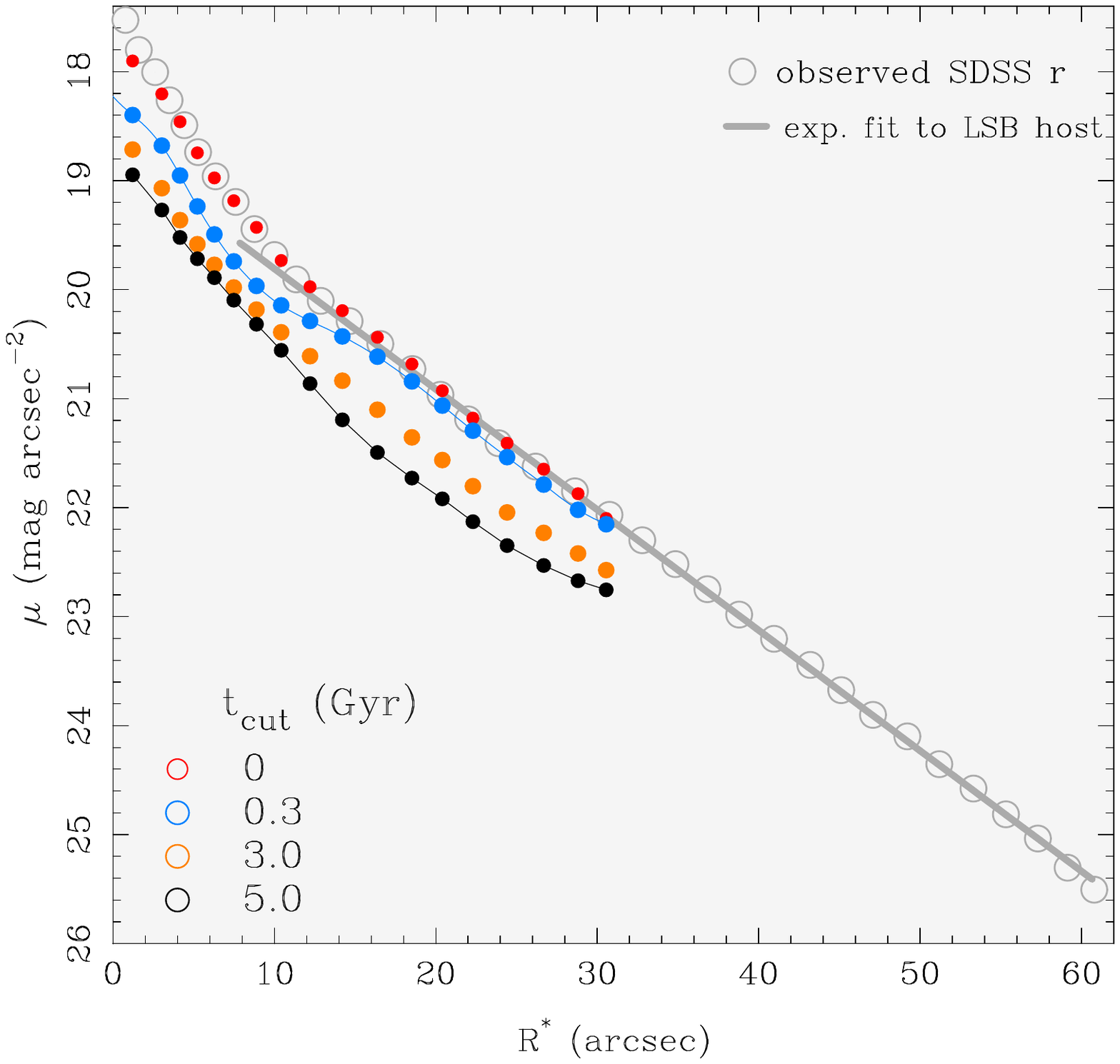}}
\put(5.0,0.36){\includegraphics[height=4.1cm, angle=0]{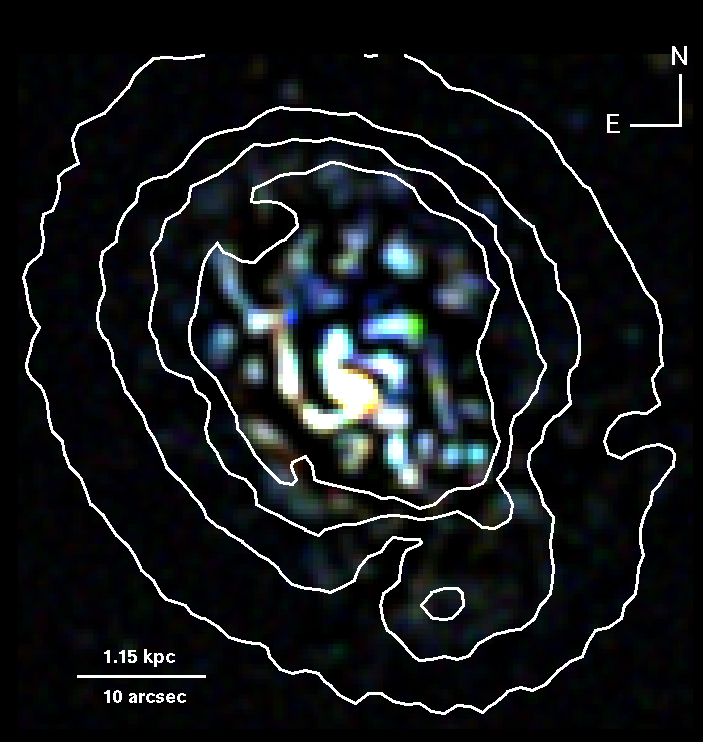}}
\PutLabel{3.8}{3.3}{\color{black}\icap a}
\PutLabel{8.57}{0.47}{\color{white}\icap b}
\end{picture}
\caption[]{\rem{a:} Surface brightness profile (SBP) of \object{NGC 7625} (open circles) computed from SDSS $r$-band data with the overlaid gray line showing a linear fit to the LSB host for \rr$\geq$20\arcsec. Filled circles show synthetic $r$-band SBPs obtained by applying \ry\ to CALIFA IFS data for four \tcut\ values (0, 0.3, 3, and 5 Gyr).
\rem{b:} true-color composite image computed by applying the flux-conserving unsharp masking technique by \citet{P98} 
to the SDSS $g$, $r,$ and $i$ images. The contours delineate the morphology of the emission-line free stellar continuum between 6390 \AA\ and 6490 \AA, as obtained from CALIFA IFS data.}
\label{RY-n7625-b}
\end{figure}

The first galaxy, \object{NGC 3391} (D=50.9 Mpc; Fig.~\ref{RY-n3991}),
hosts intense SF activity at the NE tip of an elongated
cometary (also referred to as tadpole) host galaxy (cf. the true-color SDSS image composite
in Fig.~\ref{RY-n3991}a) that manifests itself on a $g$-$i$ color
--0.1\dots\--0.5 mag and an EW(\ha)$\sim$300 \AA\ over a region of
$\sim$2.5 kpc in diameter. Spectral fits indicate that SF in that
region has been ongoing since $\sim 10^8$ yr with the young ($\leq$30
Myr $\equiv$ \tcut$_{\rm , 1}$), ionizing stellar component
representing $\approx$4\% of \mstar.  The synthetic surface brightness
$\mu$ (\sbb) maps in panels \rem{b}\&\rem{c} were computed from the
IFS data cubes by convolving the observed ($\mu_{\rm obs}$) and
modeled ($\mu_{\rm fit}$) SED with the SDSS $r$-band filter
transmission curve.  Assuming for simplicity that nebular continuum
emission is negligible \citep[e.g., however, see][]{Krueger1995,P98},
subtraction of $\mu_{\rm fit}$ from $\mu_{\rm obs}$ permits
quantification of the luminosity enhancement ($\approx$--0.2 mag,
corresponding to 17\% of the $r$-band line-of-sight intensity) due to
nebular line contamination in the NE SF region.  Panels
\rem{d}\&\rem{e} show, respectively, the reconstructed $\mu_r$ for the
stellar component older than 30 Myr and its difference to $\mu_{\rm
  obs}$. Much like in Fig.~\ref{fig:RY-spectrum}a, suppression of the
young SF component (along with nebular emission) has a striking effect
on the SED, revealing a complex $\mu_r$-enhancement pattern by
$\sim$--0.6 mag all over the NE half of \object{NGC 3991} and up to
--1.2 mag in its off-center SF knot (panel \rem{e}).  As apparent from
comparison with the \ha\ and EW(\ha) maps (panels \rem{f}\&\rem{g}),
there is a good spatial correspondence between the $\mu_r$ enhancement
because of ongoing SF (panel \rem{e}) and its nebular tracers, which adds
circumstantial support to the soundness of the approach taken.
 
A second example of the 2D application of \ry\ to IFS data is given in
Fig.~\ref{RY-n7625} for the luminous BCD \object{NGC 7625}
\citep[D=23.7 Mpc,][]{TM81,C01,C12}. Panel \rem{d} shows the $\mu_r$
map of the stellar component for $t\geq$\tcut $_{\rm , 1}$ with panel
\rem{e} revealing a complex $\mu_r$ enhancement pattern reflecting the
interplay between young stars, dust obscuration, and extended nebular
emission (cf. panels \rem{f}\&\rem{g}).

With an increasing \tcut, \ry\ uncovers a progressively smooth,
lower surface brightness (LSB) elliptical host with a central $\mu_r$
of 19 \sbb\ for a \tcut=5 Gyr.  This value is 1.3 mag fainter than
the observed $\mu$ and in excellent agreement with the results from
$R$-band profile decomposition by \citet{P98a}. The potential
  of \ry\ for surface photometry studies of star-forming galaxies can
  be better appreciated from panel \rem{a} of Fig. \ref{RY-n7625-b},
  where we show with open circles the observed SDSS $r$-band surface
  brightness profile (SBP) of \object{NGC 7625} together with
  synthetic $r$-band SBPs (solid circles) computed from the CALIFA IFS
  data cube for four \tcut\ values. It can be seen that the synthetic
  SPB for \tcut=0 Gyr matches the observed SPB out to the maximum
  photometric radius (\rr$\la$30\arcsec) corresponding to the CALIFA
  field of view, except for its innermost ($\la$3\arcsec) part because of the lower FWHM ($\sim$2\farcs7) of the IFS data. At \tcut=0.3 Gyr
  (blue symbols), the SBP suggests a flattening of the exponential
  intensity profile of the LSB host inward of $\sim$15\arcsec, which
  might be reproduced by, for example, the modified exponential fitting
  function proposed in \citet{P96a} or a S\'ersic profile with
  $\eta<1$. This pronounced intensity flattening at intermediate radii
  also implies that SF activity in \object{NGC 7625} is not confined
  to the innermost part of the LSB host, as SDSS true-color images
  suggest, but is instead spatially extended over the central
  $\sim$30\arcsec\ of the BCD. Independent support of this conjecture
  comes from unsharp masking of SDSS $g$, $r,$ and $i$ images, which
  reveals a complex pattern of multiple blue SF knots within the
  central $\sim$3.5 kpc of \object{NGC 7625} (panel \rem{b}). Finally,
  stars older than 3 Gyr and 5 Gyr (orange and black symbols,
  respectively) describe a nearly exponential distribution with a
  scale length close to that obtained from a linear fit to the
  observed SBP of the LSB host for \rr$\geq$20\arcsec\ (straight gray
  line). It is interesting to note that the central surface brightness
  implied from the latter fit to observed data ($\sim$19 \sbb) is
  consistent with the central $\mu_r$ read off the synthetic SBP for
  \tcut=5 Gyr, a fact pointing to the additional potential of \ry\ as
  a supportive tool to profile decomposition studies of star-forming
  galaxies.

\section{Discussion and summary \label{disc}}
Before providing an outline of potential applications of \ry, we
offer some cautionary notes: Evidently, the output from
\ry\ sensitively relies on the quality and soundness of solutions from
\sps\ models.  These are known to suffer from a number of
deficiencies, such as the notorious age-metallicity degeneracy, which
may propagate into hardly predictable and as yet poorly explored
biases in best-fitting SFHs.  Additionally, the rather restricted
number of library SSPs allowed by state-of-the-art \sps\ codes (at
maximum 300 for \SL) permits a rather coarse coverage of the
age and metallicity parameter space, which results in a strongly
discretized approximation to the true SFH of a galaxy. Quite
obviously, \ry\ cannot offer a better time resolution than that of
the SSP library used for the input \sps\ models. Consequently,
\ry\ can fully unfold its potential only in conjunction with
next-generation \sps\ codes that are capable of significantly alleviating
the above shortcomings.

Another aspect to bear in mind when interpreting the output from
\ry\ is that the detectability of morphological relics (e.g., past SF
episodes) does not only depend on their spectrophotometric fading, but
also on dynamical processes.  For example, differential disk rotation
acts to erase signatures of an accreted satellite within a
few rotational periods ($\sim$250 Myr for a MW-like system). Likewise,
\sstar\ enhancements due to minor mergers could gradually disperse in
the presence of processes that eventually contribute to inside-out
galaxy growth, such as bar-driven stellar migration
\citep{SellwoodBinney2002,Berentzen07,Roskar08,SB14} in normal
galaxies, diffusion of newly formed stars \citep{P02,PO12} in dwarf
galaxies, or outwardly propagating SF \citep{G16c}.

On the other hand, the briefly outlined applications of \ry\ on IFS
data (Sect.~\ref{RY-IFS}) illustrate the potential of the code in
various fields of extragalactic research.  For example, a natural
by-product of \ry\ is the removal of nebular emission and the
quantification of its effect on broadband photometry. Already, with a
conservative \tcut\ on the order of the main-sequence lifetime of
ionizing stars (10-30 Myr), \ry\ permits partial suppression of bright
SF regions from synthetic images; this opens new avenues to
structural and \rem{QMI} studies of starburst galaxies where the young
stellar component and nebular emission frequently dominate within the
optical extent. Combined with deep IFS, \ry\ may thus be regarded as
an analogous yet more powerful approach to the structural properties
of high-sSFR systems than NIR photometry. This is not only because the
latter is  expensive in terms of observational time and data
reduction effort, but also because it {\sl per se} does not permit
complete suppression of stellar populations younger than an adjustable
age cutoff.
Additionally, the $K$-band \ml\ of an instantaneously formed stellar
population varies by $\sim$2 dex within 100 Myr, depending on
metallicity and initial stellar mass function, which implies that an
accurate determination of \mstar\ requires even in NIR wavelengths
prior knowledge of the SFH.
The adjustable age cutoff allowed by \ry\ is a key advantage here,
which together with an extensive set of output quantities (e.g.,
\mstar\ and synthetic images for an unlimited number of broad- and narrowbands) suggests a broad range of applications. Examples follow.

\rem{i)} Subtraction of the SF component from BCDs and other starburst
galaxies, allowing for improved determinations of the central
intensity distribution (and, henceforth, the gravitational potential)
of the underlying host galaxy.  An unresolved question in this context
is whether the latter shows an extended flat core, which under certain
assumptions would imply a central minimum in the radial stellar
density distribution \citep[][see also Noeske et
  al. 2003]{P96a}. \rem{ii)} Spatial progression of SF activity in
galaxies: This subject encompasses several particular aspects, ranging
from the hypothesis of unidirectional SF propagation in
cometary galaxies \citep{P98,P08}, to the formation history of
multiple generations of young stellar clusters (SCs) in isolated and
interacting starburst galaxies
\citep[e.g.,][]{Ostlin03,Adamo11,Whitmore07}, to the outward propagation
of a SF front in collisional ring galaxies
\citep{ASM96,Romano2008}. Adaptive removal of SCs and/or a more
diffusely distributed young stellar substrate for a set of increasing
\tcut\ can, in principle, provide a powerful technique for the
reconstruction of SF propagation patterns and their role in galaxy
build-up.
\rem{iii)} Tidal dwarf galaxy (TDG) formation: Are these entities
forming through gas collapse within a pre-existing gravitational
potential from tidally ejected stars or out of a purely gaseous
self-gravitating component?  \citep[e.g.,][see also Duc 2012 for a
  review]{Weilbacher02}.  Subtraction of the SF component with
\ry\ could add decisive constraints for discriminating between both
scenarios.
\rem{iv)} Galaxy evolution in clusters: Galaxies falling onto clusters
may experience a complex SFH involving, for example, an initial starburst
episode followed by ram-pressure induced cessation of SF
\citep[e.g.,][]{Poggianti99} in some cases accompanied by kpc-long,
UV-emitting SF tails \citep[e.g.,][]{Hester2010,Kenney14}. \ry\ offers
a handy tool to explore galaxy evolution in clusters back to several
$10^8$ yr, i.e., over the critical phase of ram-pressure stripping.
\rem{v)} Last but not least, a natural application of \ry\ concerns
robust determinations of \rem{QMI} sets, such as CAS
\citep[concentration-asymmetry-smoothness;][]{Conselice2003} and the
Gini coefficient \citep{Lotz2004}, after decontamination of IFS data
cubes from SF or directly from \sstar\ maps.

The above examples outline the potential and wide range of
applications of \ry\ toward deciphering the galaxy assembly history in
the modern era of integral field spectroscopy.

\begin{acknowledgements}
We would like to thank the anonymous referee for valuable comments and
suggestions. JMG acknowledges support by Funda\c{c}\~{a}o para a
Ci\^{e}ncia e a Tecnologia (FCT) through the Fellowship
SFRH/BPD/66958/2009 and POPH/FSE (EC) by FEDER funding through the
Programa Operacional de Factores de Competitividade
(COMPETE). PP is supported by FCT through the Investigador FCT
Contract No. IF/01220/2013 and POPH/FSE (EC) by FEDER funding through
the program COMPETE. JMG\&PP acknowledge support by the FCT under
project FCOMP-01-0124-FEDER-029170 (Ref. PTDC/FIS-AST/3214/2012),
funded by FCT-MEC (PIDDAC) and FEDER (COMPETE) and the Exchange
Programme ``Study of Emission-Line Galaxies with Integral-Field
Spectroscopy'' (SELGIFS, FP7-PEOPLE-2013-IRSES-612701), funded by the
EU through the IRSES scheme. This paper is based on data from the
Calar Alto Legacy Integral Field Area Survey, CALIFA
(http://califa.caha.es), funded by the Spanish Ministery of Science
under grant ICTS-2009-10, and the Centro Astron\'omico
Hispano-Alem\'an.
This research made use of the NASA/IPAC Extragalactic Database (NED),
which is operated by the Jet Propulsion Laboratory, California
Institute of Technology, under contract with the National Aeronautics
and Space Administration.
\end{acknowledgements}
%

\end{document}